\documentclass[11pt]{article}

\usepackage[dvips]{geometry}
\usepackage{eurosym}
\usepackage{graphicx}
\usepackage{epstopdf}
\usepackage{amssymb,amsmath}
\usepackage{graphicx,float}
\usepackage{indentfirst}
\newcommand{\beq}{\begin{equation}}
\newcommand{\eeq}{\end{equation}}
\newcommand{\bea}{\begin{eqnarray}}
\newcommand{\eea}{\end{eqnarray}}
\newcommand{\nn}{\nonumber}
\DeclareMathOperator{\sech}{sech}

\begin{document}

\title{Multikink solutions and deformed defects}
\author{G. P. de Brito\thanks{gustavopazzini@gmail.com} and A. de Souza Dutra\thanks{dutra@feg.unesp.br} \\
\\
\textit{UNESP Univ Estadual Paulista - Campus de Guaratinguet\'a - DFQ}\\
\textit{Av. Dr. Ariberto Pereira Cunha, 333, Guaratinguet\'{a} SP Brasil}}

\maketitle

%\pacs{11.10.Kk, 11.10.Lm, 11.27.+d}

\begin{abstract}
At the present work we consider an application of the deformation procedure that enable us to construct, systematically, scalar field models supporting multikinks. We introduce a new deformation function in order to realize this task. We exemplify the procedure with three different starting models already known in the literature, and the resulting deformed models have rich vacua structures which are responsible for the appearance of multikink configurations.
\end{abstract}
\maketitle

\section{INTRODUCTION}

In the last few decades, defects structures have received considerable amount of attention in the literature. In fact, topological defects play an import role in the development in several branches of physics, from condensed matter to high energy physics and cosmology \cite{rajaraman,vilenkin,manton,vachaspati}. In field theory they usually emerge from models supporting spontaneous symmetry breaking, and they may appear as kinks, domain walls, vortices, strings and monopoles.\\
\indent In condensed matter, a recent and interesting example regarding topological defects is related with the study of magnetic domain wall in a nanowire, designed for the development of magnetic memory \cite{bischof}. Also in the context of condensed matter, in references \cite{Su,Jackiw,Chung} was shown that the presence of kink-like defects in quasi-one-dimensional systems like polyacetylene is important for the increase of conductivity to almost metallic level of this insulator, when it is introduced charged defects by doping. In high energy physics we may cite, for instance, the importance of defect structures in brane world scenarios, where we may interpret that we live in a domain-wall with 3+1 dimensions embedded in a 5-dimensional spacetime \cite{rubakov}. In the context of braneworlds with warped geometry, kink-like defects are used to engender the 5-dimensional spacetime structure \cite{gremm}. In cosmology, topological defects may be related with phase transitions in the early Universe, such defects may have been formed as the Universe cooled and various local and global symmetries were broken \cite{vilenkin,kibble}.\\
\indent Some time ago, Peyrard and Kruskal \cite{peyrard} discovered that a single kink becomes unstable when it moves in a discrete lattice with large velocities, while multikink solutions remains stable. This effect is associated with the interaction between the kink and the radiation, and the resonances were already observed experimentally \cite{kinkinteraction}. The above reasons motivated the study of multikinks and, some years ago, Champney and Kivshar \cite{champney} performed an analysis on the reasons of the appearance of multikinks in dispersive nonlinear systems. Furthermore, multikinks have applications, for instance, in the study of mobility hysteresis in a damped driven commensurable chain of atoms \cite{braun}. Moreover, in arrays of Josephson junctions, instabilities of fast kinks generate bunched fluxon states presenting multikink profiles \cite{ustinov}. However, at the present state there is a few number of scalar field models that presents multikink solutions. Recently, we proposed a model with smooth potential which supports multikink configurations \cite{multikinks}, however, in that case the ``smoothness" of the potential is guaranteed only up to the first derivative. In another paper, we proposed a model for doublets of scalar field where one of its components is a special case of multikinks, the triple-kink \cite{orbitbased}.\\
\indent Since kink-like configurations are obtained from nonlinear field equations, it is difficult to obtain analytical solutions and, as a consequence, any method that help us in this task would be certainly welcome. Some years ago, Bazeia and collaborators introduced the so-called deformation procedure \cite{DEFORMED}, which enable us to construct new scalar field models from a starting one. This procedure was wisely applied to a large number o scalar field models, see \cite{DEFORMED2} and references therein.\\
\indent In this paper we will consider an application of the deformation procedure that enable us to construct, in a systematic way, new scalar field models which supports multikink solutions. In order to realize this task, we introduce a new deformation function. The procedure will be exemplified with three different starting models that are well known in the literature, and the resulting models possess rich vacua structures which are responsible for the appearance of multikink configurations.\\
\indent This work is organized as follows: in section \ref{DEFORMATION} we review basic ideas on the deformation procedure that will be necessary for this work. In section \ref{MULTIKINKS} we introduce the deformation function that will be used in this paper and we analyze some of its properties. We also consider three examples of the generation of multikinks. In section \ref{STABILITY} we discuss the stability against small fluctuations. Finally, in section \ref{FINAL} we conclude.

\section{DEFORMATION PROCEDURE \label{DEFORMATION}}

Some years ago Bazeia and collaborators \cite{DEFORMED} introduced a procedure that enable us to construct new models supporting topological defects from a starting model. In this section we will review the main aspects on the deformation procedure. Let us consider two models of real scalar field in 1+1 dimensions described by the respective Lagrangian densities
\beq
\mathcal{L}_j = \frac{1}{2}\partial_{\mu}\phi_j \partial^{\mu}\phi_j - V_j(\phi_j) \quad , \quad
\mathcal{L}_i = \frac{1}{2}\partial_{\mu}\phi_i \partial^{\mu}\phi_i - V_i(\phi_i).
\eeq
The first order equations for the static solutions of minimal energy configuration are given by
\beq
\frac{1}{2} \bigg(\frac{d \phi_j}{dx} \bigg)^2= V_j(\phi_j) \quad \textmd{and}  \quad \frac{1}{2} \bigg(\frac{d \phi_i}{dx} \bigg)^2= V_i(\phi_i) .
\eeq
It is possible to introduce a function $\phi_j = f_{ji}(\phi_i)$, called deformation function, that connects the model described by $\mathcal{L}_i$ to the model $\mathcal{L}_j$ by relating the potentials $V_i(\phi_i)$ and $V_j(\phi_j)$ in the very specific form
\beq \label{deform_pot}
V_i(\phi_i) = \frac{V_j(f_{ij}(\phi_i))}{f'_{ij}(\phi_i)^2},
\eeq
where the prime denotes the derivative with respect to the argument of the function. This procedure allow us to start with a real scalar field model described by the Lagrangian density $\mathcal{L}_j$ whose topological solutions $\phi_j(x)$ are known, and then obtain a new real scalar field model described by $\mathcal{L}_j$, whose potential $V_i(\phi_i)$ is specified by (\ref{deform_pot}), and its topological solution $\phi_i(x)$ may be directly obtained by the inverse of the deformation function, \textit{i.e.} $\phi_i(x) = f_{ji}^{-1}(\phi_j(x))$. If $V_j(\phi_j)$ and $V_i(\phi_i)$ are positive definite, then we may express $V_j$ and $V_i$ in terms of superpotential functions $W_j(\phi_j)$ and $W_i(\phi_i)$, such that
\beq \label{superpotential}
V_j(\phi_j) = \frac{1}{2}W'_j(\phi_j)^2 \quad \textmd{and} \quad V_i(\phi_i) = \frac{1}{2}W'_i(\phi_i)^2.
\eeq
In this case, the first order equations may be written in terms of the superpotential functions as follows
\beq
\frac{d \phi_j}{dx} = W'_j(\phi_j) \quad \textmd{and} \quad \frac{d \phi_i}{dx} = W'_i(\phi_i) .
\eeq
The correspondence between the deformed superpotential with the original one, is given by
\beq \label{deform_superpot}
W'_i(\phi_i) = \frac{W'_j(f_{ji}(\phi_i))}{f'_{ij}(\phi_i)} .
\eeq
For examples of applications of the deformation procedure see \cite{DEFORMED,DEFORMED2} and references therein.

\section{MULTIKINKS FROM DEFORMED DEFECTS \label{MULTIKINKS}}

In this paper we will introduce a new kind of deformation function and explore its consequence when  applied to some models already known in the literature. We will also show that an interesting consequence arises when we apply this deformation function successive times. Let us start with a model $\mathcal{L}_j$, whose static solution is $\phi_j(x)$. We introduce a deformation function such that
\beq \label{deform_func}
\phi_j = f_{ji}(\phi_i) = \frac{1}{1 + a_i} \left(\phi_i - b + \sqrt{(\phi_i + a_i b)^2 - (1 - a_i^2)(1 - b^2)}\right) ,
\eeq
where $a_i$ and $b$ are real parameters, and, in order to ensure that $f_{ij}$ is a real function we impose that $|a_i| < 1$ and $|b|\geq 1$. The derivative with respect to $\phi_i$ is given by
\beq
\frac{d f_{ji}}{d \phi_i} = \frac{1}{1 + a_i} \left(1 + \frac{\phi_i + a_i b}{\sqrt{(\phi_i + a_i b)^2 - (1 - a_i^2)(1 - b^2)}}\right).
\eeq
As one can see, there is no finite value of $\phi_i$ such that the right side of the last equation becomes zero, thus, the deformed potential $V_i(\phi_i)$ is well defined for the above deformation function. Since we have defined $\phi_j = f_{ji}(\phi_i)$, we may write the inverse of the deformation function as follows\footnote{In fact, to define an inverse function $f_{ji}^{-1}$ to the deformation function should be necessary that $f_{ji}$ is an injetive and surjective map. In the present case we may verify that only injective condition is satisfied, and in this case we may merely ensure the existence of the left inverse of $f_{ji}$. However, we may restrict the range of the deformation function to be coincident with its image, in this case we may also ensure the existence of the right inverse, as a consequence the inverse map $f_{ji}^{-1}$ will be well defined.}
\beq \label{inverse_deform}
\phi_i = f_{ji}^{-1}(\phi_j) = \frac{1 + a_i}{2}(\phi_j + b) + \frac{(1 - a_i)(1 -b^2)}{2(\phi_j + b)} - a_i b .
\eeq

\begin{figure}[H]
\centering
\includegraphics[scale=1.3]{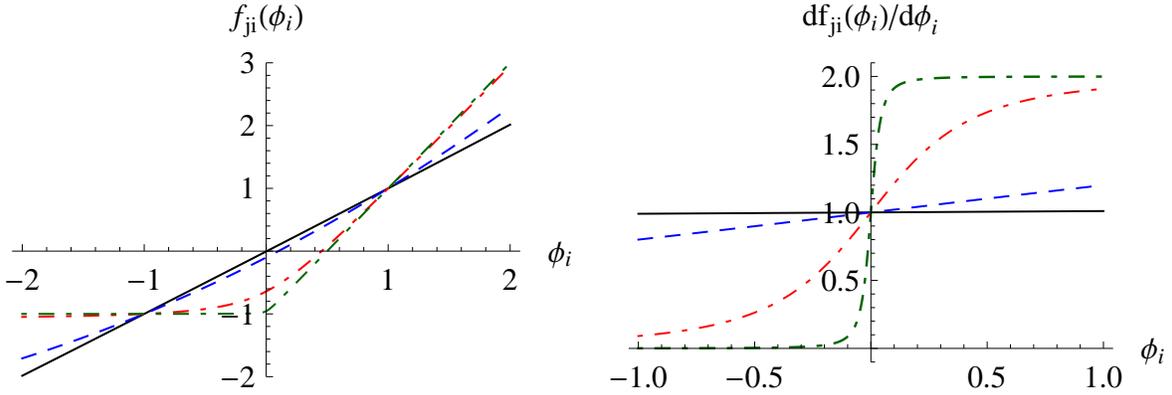}
\caption{\footnotesize{Left: Deformation function; Right: First derivative of the deformation function - Solid line (black): $b = 100$ and $a_i = 0$; Dashed line (blue): $b = 5$ and $a_i = 0$; Dot-dashed line (red): $b = 1,1$ and $a_i = 0$; Dot-dashed line (green): $b = 1,001$ and $a_i = 0$.}}
\label{FIG1}
\end{figure}

Before going further and study some explicit models, we will make some comments about general aspects of the above deformation function:
\begin{itemize}
\item Let $\tilde{\phi}_j$ be a vacuum of the original model, whose potential is given by $V_j(\phi_j)$, \textit{i.e.} $V_j(\tilde{\phi_j}) = 0$. We will demonstrate that $\tilde{\phi}_i = f_{ji}^{-1}(\tilde{\phi}_j)$ is a vacuum of the deformed model, whose potential is $V_i(\phi_i)$. In fact, from Eq. (\ref{deform_pot}) we have
\beq \label{vacua_struc}
V_i(\tilde{\phi}_i) = \frac{1}{f'_{ij}(\tilde{\phi}_i)^2} V_j(f_{ij}(\tilde{\phi}_i)) =  \frac{1}{f'_{ij}(\tilde{\phi}_i)^2} V_j[f_{ij}(f_{ji}^{-1}(\tilde{\phi}_j))] = \frac{1}{f'_{ij}(\tilde{\phi}_i)^2} V_j(\tilde{\phi}_j) = 0,
\eeq
since $V_j(\tilde{\phi}_j) = 0$. Therefore, $\tilde{\phi}_i = f_{ji}^{-1}(\tilde{\phi}_j)$ is a vacuum of the deformed model. Note that the above argumentation is independent of the explicit form of the deformation function.

\item Since we are interested in topological solutions, we may look to the behaviour of the deformation function $f_{ji}$ in a determined topological sector of the deformed model, for instance, in the region $\phi_i \in [\tilde{\phi}_i,\bar{\phi}_i]$, where $\tilde{\phi}_i = f_{ji}^{-1}(\tilde{\phi}_j)$ and $\bar{\phi}_i = f_{ji}^{-1}(\bar{\phi}_j)$ are vacua of the model $V_i(\phi_i)$. Let us investigate the behaviour of $f_{ji}$ in this sector, considering that $b$ assumes large values compared to $\phi_i$. It is not difficult to see that in this condition we have
\beq \nonumber
\sqrt{(\phi_i + a_i b)^2 - (1 - a_i^2)(1 - b^2)} \approx b + a_i \phi_i .
\eeq
Thus, substituting the last expression in Eq. (\ref{deform_func}) we obtain $f_{ji}(\phi_i) \approx \phi_i$, \textit{i.e.}, for large values of $b$ the deformation function behaves as in identity map, see FIG. \ref{FIG1}.

\item Now, we investigate the behaviour of $f_{ji}(\phi_i)$ for the case $b = 1 + \varepsilon$, where $\varepsilon$ is a very small and positive parameter. In this case we have $1 - b^2 \approx 0$, and, as a consequence
\beq \nn
\sqrt{(\phi_i + a_i b)^2 - (1 - a_i^2)(1 - b^2)} \approx |\phi_i + a_i b |.
\eeq
Substituting this result in Eq. (\ref{deform_func}), we obtain
\beq
f_{ji}(\phi_i) \approx \frac{\phi_i - b + |\phi_i + a_i b|}{1 + a_i} = \begin{cases}
\frac{2}{1+a_i}\phi_i - \frac{1 - a_i}{1 + a_i}b \,, \quad \phi_i \geq -a_i b  \\
-b \, , \quad \phi_i \leq -a_i b
\end{cases}.
\eeq
As one can see in FIG. \ref{FIG1} the dot-dashed-green line is in accordance with the above expression.
\end{itemize}

Below, we will consider three applications of the deformation function above introduced and explore the possibility of obtaining multikink solutions.

\subsection{Deformed model I}

\begin{figure}[H]
\centering
\includegraphics[scale=1.3]{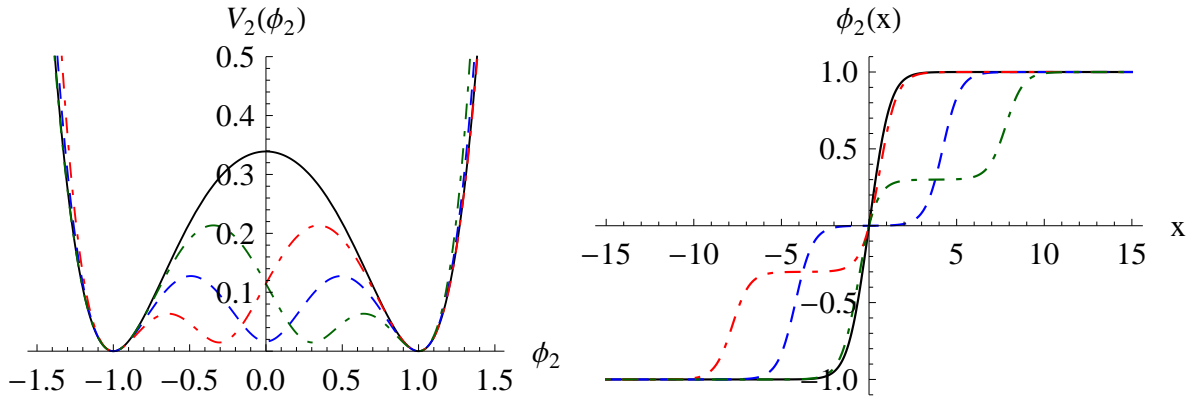}
\caption{\footnotesize{Left - Solid line (black): $b = 1,4$ and $a_2 = 0$; Dashed line (blue): $b = 1,005$ and $a_2 = 0$; Dot-dashed line (red): $b = 1,005$ and $a_2 = 0,3$; Dot-dashed line (green): $b = 1,005$ and $a_2 = -0,3$. Right - Solid line (black): $b = 1,4$ and $a_2 = 0$; Dashed line (blue): $b = 1,0000001$ and $a_2 = 0$; Dot-dashed line (red): $b = 1,0000001$ and $a_2 = 0,3$; Dot-dashed line (green): $b = 1,0000001$ and $a_2 = -0,3$.}}
\label{FIG2}
\end{figure}

We will first consider the application of the above deformation function in the so called $\phi^4$ model, which potential function is given by
\beq \label{Pot_1_phi4}
V_1(\phi_1) = \frac{1}{2} (1 - \phi_1^2)^2,
\eeq
the correspondent superpotential is given by
\beq
W_1(\phi_1) = \phi_1 - \frac{1}{3}\phi_1^3 .
\eeq
The usual kink solution is given by \footnote{It is well known that this model also supports anti-kink solutions, however, we do not take this in consideration here. But, we emphasize that the same analysis performed in this section may also be applied to anti-kink solutions.}
\beq
\phi_1(x) = \tanh(x - x_0).
\eeq
Where $x_0$ is an integration constant that determines the center of the kink. We may introduce a deformed model $\mathcal{L}_2$ using the deformation function given by Eq. (\ref{deform_func})
\beq \nn
\phi_1 = f_{1,2}(\phi_2) = \frac{1}{1 + a_2} \left(\phi_2 - b + \sqrt{(\phi_2 + a_2 b)^2 - (1 - a_2^2)(1 - b^2)}\right) .
\eeq
Using Eq. (\ref{deform_pot}) we obtain the deformed potential $V_2(\phi_2)$
\bea \label{Pot_2_phi4}
V_2(\phi_2) = \frac{(1 + a_2)^2\left((\phi_2 + a_2 b)^2 - (1 - a_2^2)(1 - b^2)\right)}{2 \left(\phi_2 + a_2 b + \sqrt{(\phi_2 + a_2 b)^2 - (1 - a_2^2)(1 - b^2)}\right)^2} \times \nn \\ \times \left[1 - \left( \frac{1}{1 + a_2} \left(\phi_2 - b + \sqrt{(\phi_2 + a_2 b)^2 - (1 - a_2^2)(1 - b^2)}\right) \right)^2 \right]^2.
\eea
In terms of the superpotential we obtain
\bea
&W'_2(\phi_2)& = \frac{(1 + a_2)\sqrt{(\phi_2 + a_2 b)^2 - (1 - a_2^2)(1 - b^2)}}{\phi_2 + a_2 b + \sqrt{(\phi_2 + a_2 b)^2 - (1 - a_2^2)(1 - b^2)}} \times \nn \\ &\times & \left[1 - \left( \frac{1}{1 + a_2} \left(\phi_2 - b + \sqrt{(\phi_2 + a_2 b)^2 - (1 - a_2^2)(1 - b^2)}\right) \right)^2 \right].
\eea
It is interesting to note that despite of the non-usual expression of this potential, its graphical behaviour is quite usual. As one can see in FIG. \ref{FIG2} (left), the potential $V_2(\phi_2)$ has two global minima localized at $\phi_2 = \pm 1$, and also, there is a local minimum when the parameter $b$ is close to the critical value $b=1$. Also, we can see that the parameter $a_2$ controls the symmetry of the potential. It is interesting to note that the symmetrical case ($a_2 = 0$) reproduces the same kind of potential considered in ref. \cite{hott}, which has interesting implications in braneworld scenarios.

\begin{figure}[H]
\centering
\includegraphics[scale=1.3]{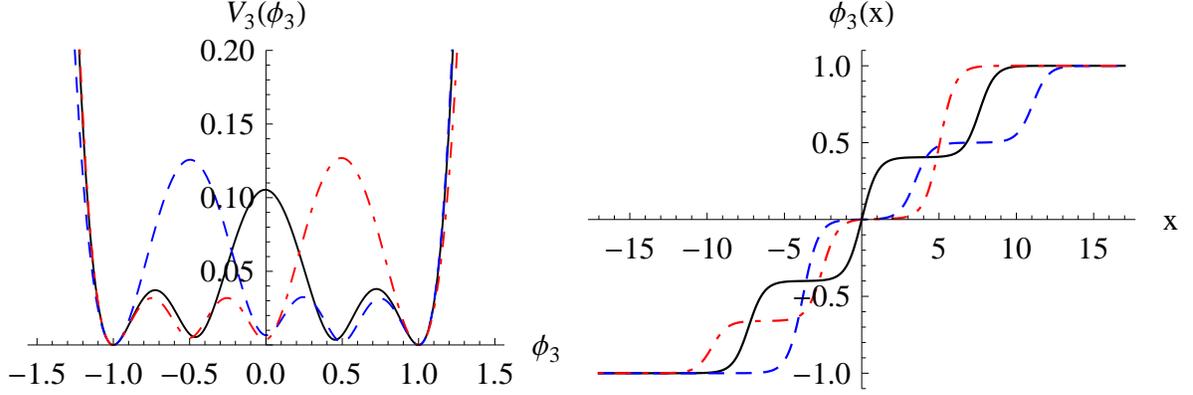}
\caption{\footnotesize{Left - $b = 1,002$; Solid line (black): $a_2 = -0,25$  and $a_3 = 0,46$; Dashed line (blue): $a_2 = 0$  and $a_3 = 0$; Dot-dashed line (red): $a_2 = 0,34$  and $a_3 = 0,5$. Right - $b = 1,0000005$; Solid line (black): $a_2 = -0,15$ and $a_3 = 0,4$; Dashed line (blue): $a_2 = 0$ and $a_3 = 0$; Dot-dashed line (red): $a_2 = 0,2$ and $a_3 = 0,66$.}}
\label{FIG3}
\end{figure}

The topological solution for this model may be obtained by using the inverse deformation function given by Eq. (\ref{inverse_deform}), namely
\beq \label{Solution_2_phi4}
\phi_2(x) = f_{1,2}^{-1}(\phi_1(x)) = \frac{1 + a_2}{2}(\tanh(x - x_0) + b) + \frac{(1 - a_2)(1 -b^2)}{2(\tanh(x - x_0) + b)} - a_2 b .
\eeq
As one can see in FIG. \ref{FIG2} (right) the profile of the field $\phi_2(x)$ corresponds to a 2-kink solution. Note that the parameter $b$ controls the wide of the additional ``step" that arises, furthermore, the parameter $a_2$ engender an asymmetrical behaviour to the 2-kink solution. We emphasize that the above solution appear in a model with two real scalar fields in the context of braneworlds \cite{Asymm_BNRT}, in that case the asymmetrical behaviour combined with the appearance of double domain wall solutions play an important role when addressing the hierarchy problem in thick brane scenarios.\\
\indent Now, let us turn the attention on how to obtain multikink solutions. We will proceed with this by successive applications of the deformation procedure. Now, let $\mathcal{L}_2$ be our original model, we will construct a new model $\mathcal{L}_3$ applying the deformation procedure to $\mathcal{L}_2$ with deformation function $\phi_2 = f_{2,3}(\phi_3)$ given by
\beq
\phi_2 = f_{2,3}(\phi_3) = \frac{1}{1 + a_3} \left(\phi_3 - b + \sqrt{(\phi_3 + a_3 b)^2 - (1 - a_3^2)(1 - b^2)}\right) .
\eeq
Using the last expression in Eq. (\ref{deform_pot}) we may obtain the new deformed potential $V_3(\phi_3)$
\bea \label{Pot_3_phi4}
&V_3(\phi_3)& = \frac{(1 + a_3)^2\left( (\phi_3 + a_3 b)^2 - (1 - a_3^2)(1 - b^2)\right)}{\left(\phi_3 + a_3 b + \sqrt{(\phi_3 + a_3 b)^2 - (1 - a_3^2)(1 - b^2)}\right)^2} \nn \times \\ &\times & \frac{(1 + a_2)^2\left((f_{2,3}(\phi_3) + a_2 b)^2 - (1 - a_2^2)(1 - b^2)\right)}{2 \left(f_{2,3}(\phi_3) + a_2 b + \sqrt{(f_{2,3}(\phi_3) + a_2 b)^2 - (1 - a_2^2)(1 - b^2)}\right)^2} \times \nn \\ &\times & \left[1 - \left( \frac{1}{1 + a_2} \left(f_{2,3}(\phi_3) - b + \sqrt{(f_{2,3}(\phi_3) + a_2 b)^2 - (1 - a_2^2)(1 - b^2)}\right) \right)^2 \right]^2 .
\eea
Note that in the above equation we still left some terms with dependence in the deformation function $f_{2,3}(\phi_3)$ in order to economize space. In FIG. \ref{FIG3} (left) we plot the potential $V_3(\phi_3)$ for some values of the parameters $b$, $a_2$ and $a_3$. Note that in this potential, there are two global minima localized at $\phi_3 = \pm 1$, and also, we may observe the appearance of two local minima if the parameter $b$ is close to the critical value $b=1$. Once again, the parameters $a_2$ and $a_3$ are responsible for the symmetry of the potential.\\
\indent The analytical expression for the field solution $\phi_3(x)$ may be directly obtained by the inverse deformation function $\phi_3(x) = f_{2,3}^{-1} (\phi_2(x))$, in this case we have obtained
\bea \label{Solution_3_phi4}
&&\phi_3(x) = f_{2,3}^{-1} (\phi_2(x)) = \frac{1 + a_3}{2}\left( \frac{1 + a_2}{2}(\tanh(x - x_0) + b) + \frac{(1 - a_2)(1 -b^2)}{2(\tanh(x - x_0) + b)} - a_2 b + b \right) + \nn \\ &&+\frac{(1 - a_3)(1 -b^2)}{2}\left( \frac{1 + a_2}{2}(\tanh(x - x_0) + b) + \frac{(1 - a_2)(1 -b^2)}{2(\tanh(x - x_0) + b)} - a_2 b + b \right)^{-1} - a_3 b .
\eea
In FIG. \ref{FIG3} (right) we plot $\phi_3(x)$, and as one can see, the appearance of two additional local minima in the potential engenders a 3-kink configuration. As we may expect, the parameter $b$ controls the wide of the additional ``steps", while $a_2$ and $a_3$ controls its symmetry.

\begin{figure}[H]
\centering
\includegraphics[scale=1.3]{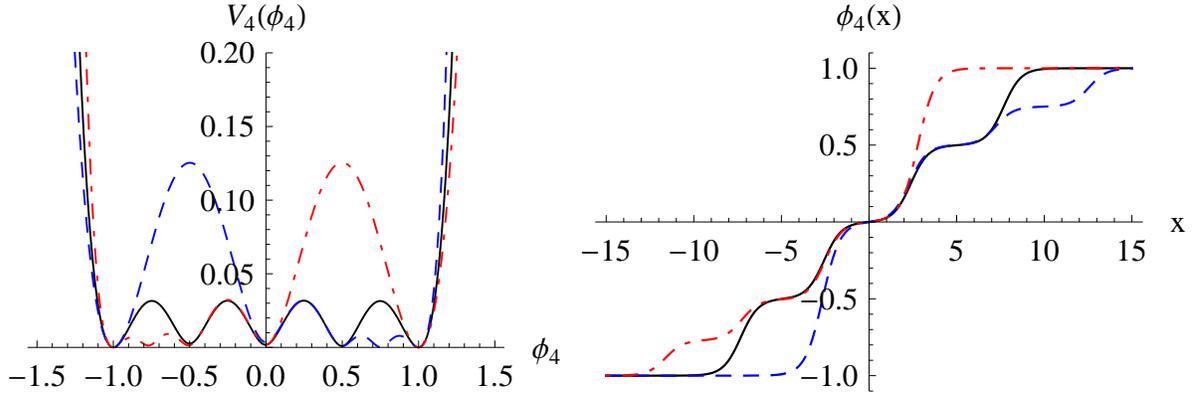}
\caption{\footnotesize{Left - $b = 1,001$; Solid line (black): $a_2 = 0$, $a_3 = 0,335$ and $a_4 = 0,5$; Dashed line (blue): $a_2 = 0$, $a_3 = 0$ and $a_4 = 0$; Dot-dashed line (red): $a_2 = 0,33$, $a_3 = 0,7$ and $a_4 = 0,77$. Right - $b = 1,00005$; Solid line (black): $a_2 = 0$, $a_3 = 0,335$ and $a_4 = 0,5$; Dashed line (blue): $a_2 = 0$, $a_3 = 0$ and $a_4 = 0$; Dot-dashed line (red): $a_2 = 0,33$, $a_3 = 0,7$ and $a_4 = 0,77$.}}
\label{FIG4}
\end{figure}

We may continue the process in a systematic way and define new deformed models from the deformation functions $\phi_3 = f_{3,4}(\phi_4)$, $\phi_4 = f_{4,5}(\phi_5)$, $\phi_5 = f_{5,6}(\phi_6)$, etc; and as a consequence we obtain its respective deformed potentials $V_4(\phi_4)$, $V_5(\phi_5)$, $V_6(\phi_6)$, etc. In FIG. \ref{FIG4} we plot the potential $V_4(\phi_4)$ and the corresponding solution $\phi_4(x)$. As one can see in FIG. \ref{FIG4} (left), when the parameter $b$ is close to the critical value $b=1$ there are three local minima in the potential, and as a consequence, in FIG. \ref{FIG4} (right) we may observe the appearance of a 4-kink configuration.

\begin{figure}[H]
\centering
\includegraphics[scale=.8]{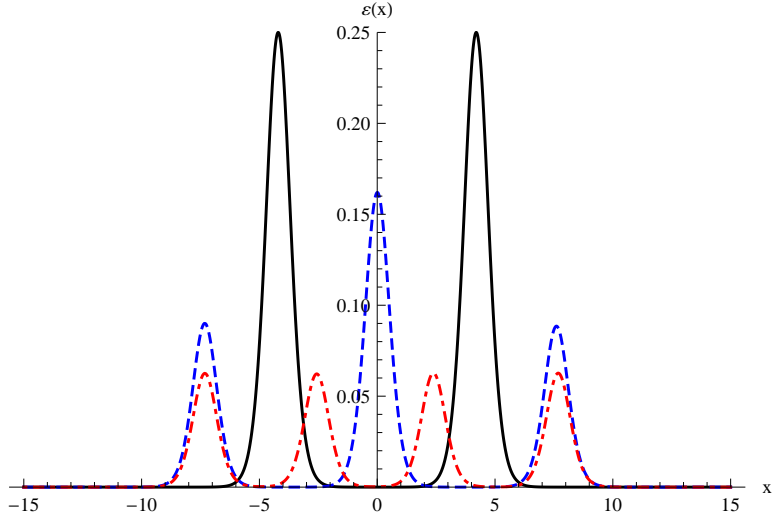}
\caption{\footnotesize{Solid line (black): energy density corresponding to $V_2(\phi_2)$ with parameters given by $b = 1,0000001$ and $a_2 = 0$. Dashed line (blue): energy density corresponding to $V_3(\phi_3)$ with parameters given by $b = 1,0000005$, $a_2 = 0$, $a_3 = 0,335$. Dot-dashed line (red): energy density corresponding to $V_4(\phi_4)$ with parameters given by $b = 1,00005$, $a_2 = 0$, $a_3 = 0,335$ and $a_4 = 0,5$.}}
\label{FIG5}
\end{figure}

Summing up: we have started with the $\phi^4$ model, which has two global minima and no local one, and, using the deformation procedure we arrive at a new model with local minima. As one can observe, each application of the deformation function (\ref{deform_func}) result in a new potential with an additional local minima, and as a consequence, the number of ``steps" in the kink-like configurations increases with successive deformations, allowing the existence of multikink configurations. \\
\indent Before we go further and consider another example, let us make some comments about he energy density of the configurations considered above. An interesting feature related with topological configurations, such as kinks, is the fact that those solutions has localized energy in space. In order to study the energy distribution for a given defect, \textit{e.g.} $\phi_j(x)$, we consider the so called energy density, namely
\beq
\varepsilon_j(x) = \frac{1}{2}\left(\frac{d \phi_j}{dx}\right)^2 + V_j(\phi_j(x)).
\eeq
Applying the deformation procedure we obtain that the energy density associated with the deformed defect, \textit{e.g.} $\phi_i(x)$, is given by
\beq
\varepsilon_i(x) = \frac{1}{g_{ji}(x)^2}\varepsilon_j(x), \quad \textmd{where} \quad g_{ji}(x) = \frac{d f_{ji}}{d\phi_i} \bigg|_{\phi_i(x)}.
\eeq
In FIG. \ref{FIG5} we plot the energy density corresponding for the three models constructed in this section. As one can see in that figure, each additional ``step" engenders a new ``peak" in the energy distribution, according with our expectation.

\subsection{Deformed model II}

\begin{figure}[H]
\centering
\includegraphics[scale=1.3]{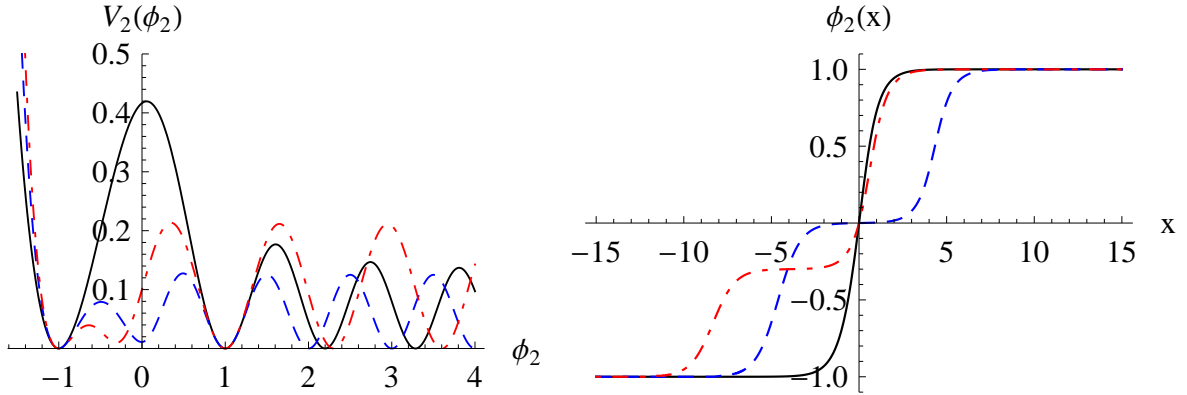}
\caption{\footnotesize{Left - Solid line (black): $b = 2$ and $a_2 = 0$; Dashed line (blue): $b = 1,005$ and $a_2 = 0$; Dot-dashed line (red): $b = 1,005$ and $a_2 = 0,3$. Right - Solid line (black): $b = 2$ and $a_2 = 0$; Dashed line (blue): $b = 1,000001$ and $a_2 = 0$; Dot-dashed line (red): $b = 1,000001$ and $a_2 = 0,3$.}}
\label{FIG6}
\end{figure}

In this second example, we will consider the application of the deformation function (\ref{deform_func}) in the so called sine-Gordon model, which potential is given by
\beq \label{Pot_1_sG}
V_1(\phi_1) = \frac{1}{2}\cos^2 \left( \frac{\pi}{2} \phi_1 \right).
\eeq
For recent applications of the deformation procedure in sine-Gordon model see ref. \cite{DEFORMED3}. The above potential has infinite degenerate minima localized at $\tilde{\phi}_1 = 1 + 2n$ where $n = 0, \pm 1, \pm 2, \pm 3,...$. Each pair of neighbour minima defines a topological sector in this model, and we will label each sector by $n$. The topological solutions of this model are given by the sine-Gordon kinks\footnote{Once again, anti-kink solutions also exist, however, we will not consider it in this paper.}
\beq
\phi_1(x) = \frac{2}{\pi} \arcsin\left[\tanh\left(\frac{\pi}{2}(x-x_0) \right) \right] + 2 n; \quad n = 0, \pm 1, \pm 2, \pm 3,....
\eeq
Note that in the above equation we have a family of kinks corresponding to different topological sectors, however, in this paper we will focus our attention only on the topological sector $n=0$, which will be the only one where multikink structures appear\footnote{Of course that we may apply the deformation function to others sectors, however, our analysis has shown that for $n \neq 0$ kink solutions are mapped into single kinks.}.\\
\indent Applying the same procedure as the previous section, one arrive at the following deformed potential
\bea \label{Pot_2_sG}
V_2(\phi_2) = \frac{(1 + a_2)^2\left((\phi_2 + a_2 b)^2 - (1 - a_2^2)(1 - b^2)\right)}{2 \left(\phi_2 + a_2 b + \sqrt{(\phi_2 + a_2 b)^2 - (1 - a_2^2)(1 - b^2)}\right)^2} \times \nn \\
\times \cos^2 \left[ \frac{\pi}{2(1 + a_2)} \left(\phi_2 - b + \sqrt{(\phi_2 + a_2 b)^2 - (1 - a_2^2)(1 - b^2)}\right) \right].
\eea
In FIG. \ref{FIG6} (left) we plot the potential $V_2(\phi_2)$ for some values of the parameters $b$ and $a_2$. As one can see in that figure, the deformed potential is quit different from the original one, note that the oscillating behaviour in $V_2$ only occurs for $\phi_2 > 0$. We may also see that all vacua of the deformed potential $V(\phi_2)$ are located ate $\phi_2 > -1$. We may observe the appearance of a local minima in the topological sector $-1 < \phi_2 < +1$. However, there is no local minima in other topological sectors, which justify our statement that the topological sector corresponding to $n=0$ is the only one in which we expect to find multikink structures. \\
\indent Using the inverse deformation function (\ref{inverse_deform}) we may obtain the explicit solution of the deformed defect, namely
\bea \label{Solution_2_sG}
\phi_2(x) = f_{1,2}^{-1}(\phi_1(x)) = \frac{1 + a_2}{2}\left( \frac{2}{\pi} \arcsin\left[\tanh\left(\frac{\pi}{2}(x-x_0) \right) \right] + b \right) + \nn \\ + \frac{(1 - a_2)(1 -b^2)}{2} \left( \frac{2}{\pi} \arcsin\left[\tanh\left(\frac{\pi}{2}(x-x_0) \right) \right] + b \right)^{-1} - a_2 b.
\eea
In FIG. \ref{FIG6} (right) we plot the solution $\phi_2(x)$ for some values of the parameters $b$ and $a_2$. As we may observe, the topological solution $\phi_2(x)$ exhibits a 2-kink profile for the case in which the parameter $b$ is close to the critical value $b=1$.

\begin{figure}[H]
\centering
\includegraphics[scale=1.3]{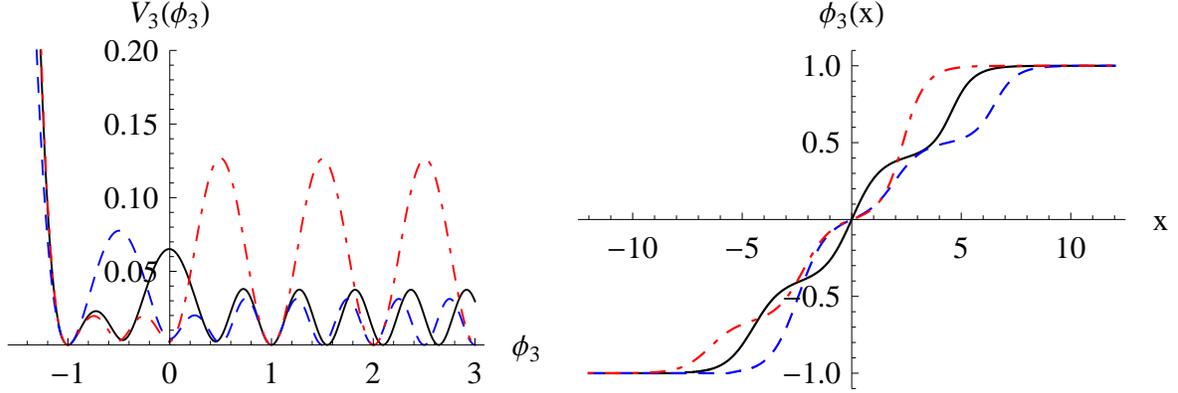}
\caption{\footnotesize{Left - $b = 1,002$; Solid line (black): $a_2 = -0,25$  and $a_3 = 0,46$; Dashed line (blue): $a_2 = 0$  and $a_3 = 0$; Dot-dashed line (red): $a_2 = 0,34$  and $a_3 = 0,5$. Right - $b = 1,001$; Solid line (black): $a_2 = -0,15$ and $a_3 = 0,4$; Dashed line (blue): $a_2 = 0$ and $a_3 = 0$; Dot-dashed line (red): $a_2 = 0,2$ and $a_3 = 0,66$.}}
\label{FIG7}
\end{figure}

Proceeding in the same way that we did in the case of the model I, we may apply the deformation function successively in order to obtain multikink structures. For instance, we may define a deformation function by $\phi_2 = f_{2,3}(\phi_3)$ , and then, we obtain a deformed potential and a deformed defect, respectively by
\beq \nn
V_3(\phi_3) = \frac{V_2 (f_{2,3}(\phi_3) )}{f_{2,3}'(\phi_3)^2} \quad \textmd{and} \quad \phi_3(x) = f^{-1}_{2,3}(\phi_2(x)).
\eeq
We emphasize that we will not write the explicit expression for $V_3(\phi_3)$ and $\phi_3(x)$ in order to economize space. In FIG. \ref{FIG7} we plot the deformed potential $V_3(\phi_3)$ and the deformed defect $\phi_3(x)$. In this case, we may observe that the deformed potential has two local minima in the topological sector $-1 < \phi_3 +1$, and as we may expect, the corresponding topological defect process a 3-kink structure.

The results presented in this section are very similar to those results discussed for the case of model I, \textit{i.e.}, each application of the deformation procedure engender an additional local minima to the potential and an additional ``step" in the topological solution, allowing multikinks configurations.

\subsection{Deformed model III}

\begin{figure}[H]
\centering
\includegraphics[scale=1.3]{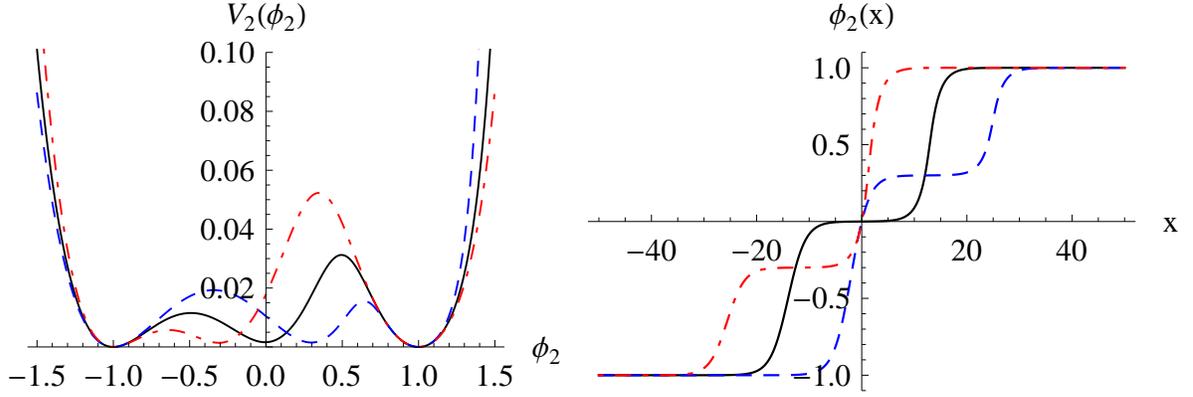}
\caption{\footnotesize{Left - $b = 1,005$; Solid line (black): $a_2 = 0$; Dashed line (blue): $a_2 = -0,3$; Dot-dashed line (red): $a_2 = 0,3$. Right - $b = 1,0000001$; Solid line (black): $a_2 = 0$; Dashed line (blue): $a_2 = -0,3$; Dot-dashed line (red): $a_2 = 0,3$.}}
\label{FIG8}
\end{figure}

\begin{figure}[H]
\centering
\includegraphics[scale=1.3]{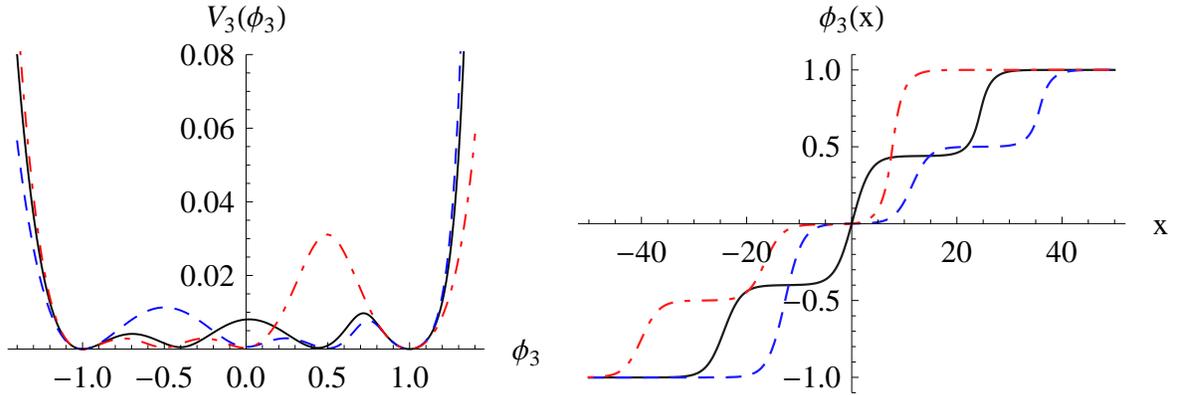}
\caption{\footnotesize{Left - $b = 1,002$; Solid line (black): $a_2 = -0,2$  and $a_3 = 0,4$; Dashed line (blue): $a_2 = 0$  and $a_3 = 0$; Dot-dashed line (red): $a_2 = 0,34$  and $a_3 = 0,5$. Right - $b = 1,0000005$; Solid line (black): $a_2 = -0,2$  and $a_3 = 0,4$; Dashed line (blue): $a_2 = 0$  and $a_3 = 0$; Dot-dashed line (red): $a_2 = 0,34$  and $a_3 = 0,5$.}}
\label{FIG9}
\end{figure}

\begin{figure}[H]
\centering
\includegraphics[scale=1.3]{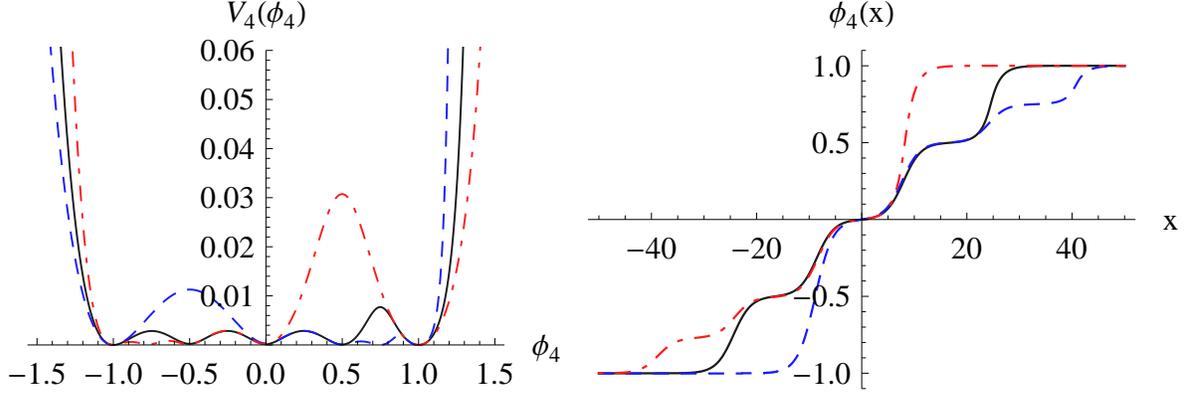}
\caption{\footnotesize{Left - $b = 1,001$; Solid line (black): $a_2 = 0$, $a_3 = 0,335$ and $a_4 = 0,5$; Dashed line (blue): $a_2 = 0$, $a_3 = 0$ and $a_4 = 0$; Dot-dashed line (red): $a_2 = 0,33$, $a_3 = 0,7$ and $a_4 = 0,77$. Right - $b = 1,00005$; Solid line (black): $a_2 = 0$, $a_3 = 0,335$ and $a_4 = 0,5$; Dashed line (blue): $a_2 = 0$, $a_3 = 0$ and $a_4 = 0$; Dot-dashed line (red): $a_2 = 0,33$, $a_3 = 0,7$ and $a_4 = 0,77$.}}
\label{FIG10}
\end{figure}

The third model to be considered in this paper is given by the following potential
\beq
V_1(\phi_1) = \frac{1}{2 \alpha^2} \left( A \cosh(\alpha \phi_1) - \sech(\alpha \phi_1) \right)^2 .
\eeq
where we have defined $\alpha = \cosh^{-1}(1/\sqrt{A})$ and we use the restriction $0<A<1$. The above potential was introduced in ref. \cite{VLdutra} to study the emergence of vacuumless system from vacuum ones. This potential has a topological sector defined by the region between the vacua $\phi_1 = -1$ and $\phi_1 = +1$, and as a consequence, we may find topological solutions in this sector, namely
\beq
\phi_1(x) = \sinh^{-1}\left( \sqrt{\frac{1 - A}{A}} \tanh(\sqrt{A(1-A)}x) \right).
\eeq
The above topological solution corresponds to a single kink connecting the vacua states $\phi_1 = -1$ and $\phi_1 = +1$.\\
\indent Instead of considering the analytical expressions for the deformed potentials and its solutions, which are considerably complicated, we will analyze only its graphical behaviour. In FIG. \ref{FIG8} we plot the deformed potential $V_2(\phi_2)$ and it solution $\phi_2(x)$ for some values of the parameters $a_2$ and $b$.
As one can see, the behaviour of this potential and its corresponding solutions are very similar to those results presented in the case of model I, the parameter $a_2$ controls the symmetry of the problem, while the parameter $b$ engender the appearance of an additional ``step" in the kink solution. For completeness, we plot in FIG. \ref{FIG9} we plot the deformed potential $V_3(\phi_3)$ and the deformed solution $\phi_3(x)$, while in FIG. \ref{FIG10} we plot the deformed potential $V_4(\phi_4)$ and the solution $\phi_4(x)$.

The result is, then, exhaustive: the parameter $b$ controls the appearance and the wide of additional ``steps" in the topological solutions, while the parameters $a_2$, $a_3$, $a_4$ controls the symmetry of the problem.

\section{STABILITY \label{STABILITY}}

One of the reasons that makes topological defects be interesting is the stability against linear fluctuations. In this section we will verify, for completeness, that those topological defects introduced in this paper are stable. The analysis of the stability is quit usual \cite{DEFORMED,DEFORMED2,JACKIW}, one consider small fluctuations around the classical field solutions and we arrive at quantum mechanical-like problem with Hamiltonian operator given by
\beq
\hat{H}_i = -\frac{d^2}{dx^2} + U_i(x),
\eeq
where
\beq \nn
U_i(x) = \frac{d^2 V_i}{d \phi_i^2} \bigg|_{\bar{\phi}_i(x)},
\eeq
and $\bar{\phi}_i(x)$ stands for the classical configuration. The investigation goes as follows: we say that the system is stable if there is no negative eigenvalue of the above Hamiltonian, otherwise, we say that the system is unstable. For those cases in which is possible to express the potential $V_i$ in terms of the so called superpotential function $W_i$, see Eq. (\ref{superpotential}), the Hamiltonian operator may be factorized in the following way
\beq
\hat{H}_i = \hat{S}_i^\dagger \hat{S}_i = \left(\frac{d}{dx} + u_i(x) \right)\left(-\frac{d}{dx} + u_i(x)\right),
\eeq
where
\beq \nn
u_i(x) = \frac{d^2 W_i}{d \phi_i^2} \bigg|_{\bar{\phi}_i(x)}
\eeq
It is not difficult to see that the operator $\hat{H}_i = \hat{S}_i^\dagger \hat{S}_i$ has only non negative eigenvalues, which ensures the stability of the systems considered in this paper\footnote{We note that, although we haven't mentioned this explicitly in the previous sections, all the models considered in this paper have positive potentials, and as a consequence, we may express them in terms of superpotentials.}. \\
\indent Up to now we have not mentioned any features of deformed defects in our analysis of stability, since we have proved the stability with general arguments. However, we will show that there exist an interesting relation between the zero mode of the deformed defect and the zero mode of the original model. Let us consider $\mathcal{L}_j$ as our original model, and $\mathcal{L}_i$ being the deformed one. The zero mode of the original model $\mathcal{L}_j$ may be determined by $\hat{S}_j \eta_j^{(0)}(x) = 0$, which solution is given by
\beq
\eta_j^{(0)}(x) = \eta_j^{(0)}(x_0)  \exp\left( \int_{x_0}^x dy \, u_j(y) \right).
\eeq
On the other hand, the zero mode of the deformed defect, which satisfies $\hat{S}_i \eta_i^{(0)}(x) = 0$, is given by
\beq \label{zero_mode}
\eta_i^{(0)}(x) = \eta_i^{(0)}(x_0)  \exp\left( \int_{x_0}^x dy \, u_i(y) \right).
\eeq
Note that we may use the correspondence between the deformed and the original superpotential, Eq. (\ref{deform_superpot}), in order to obtain
\beq
u_i(x) = u_j(x) - \frac{f''_{ji}}{{f'_{ji}}^2} \frac{d W_j(f_{ji}(\phi_i))}{d \phi_j} \bigg|_{\bar{\phi}_i(x)} .
\eeq
substituting the last one in (\ref{zero_mode}), we arrive at
\beq \label{zero_mode_2}
\eta_i^{(0)}(x) = h_{ji}(x) \eta_j^{(0)}(x),
\eeq
where we have defined
\beq
h_{ji}(x) = \frac{\eta_i(x_0)}{\eta_j(x_0)} \exp\left( - \int_{x_0}^x dx \, \frac{f''_{ji}}{{f'_{ji}}^2} \frac{d W_j(f_{ji}(\phi_i))}{d \phi_j} \bigg|_{\bar{\phi}_i(x)} \right) .
\eeq

As one can see in Eq. (\ref{zero_mode_2}), if we know the zero mode of the original model, then we may obtain the zero mode of the deformed model by multiplying the original zero mode by the function $h_{ji}(x)$. Unfortunately, Eq. (\ref{zero_mode_2}) holds only for the zero modes, and not for the excited modes. In this vein, as far we know, there is no proposal in the literature that complies this task.

\section{FINAL REMARKS \label{FINAL}}

At the present work we consider an application of the deformation procedure to generate new scalar field models supporting multikink configurations. We introduced a new deformation function which lead to interesting features when applied to usual models already considered in the literature. We performed three examples of the systematic procedure developed here. It is possible to observe with these examples that the deformation function possess a parameter that controls de symmetry of the deformed potential and there is a parameter that controls the appearance of multikinks. As one can see through the examples, those deformed potentials that possess multikink solutions have intricate analytical expressions, however, their graphical behavior is quite usual. We also considered the behavior of the energy density distribution, resulting in the appearance of space regions where the energy is concentrated, evidencing the multi-domain-wall character of the configuration. Finally, the stability analysis has shown that the defects considered in this paper are stable, as we have expected, since they are topological configurations. We also have shown that there exists a connection between the zero mode of the original model and the zero mode of the deformed one.
\bigskip

\textbf{Acknowledgements: }The authors thanks to CNPq and FAPESP for partial financial support.

\end{document}